\pdfoutput=1

\documentclass[11pt]{article}

\usepackage[preprint]{acl}

\usepackage{times}
\usepackage{latexsym}
\usepackage[T1]{fontenc}
\usepackage[utf8]{inputenc}
\usepackage{microtype}
\usepackage{inconsolata}
\usepackage{graphicx}
\usepackage{soul}
\usepackage{quoting}
\quotingsetup{vskip=0pt}

\newcommand{\gemma}{Gemma-2b}
\newcommand{\phimini}{Phi-3-mini-4k}
\newcommand{\metallama}{Meta-Llama-3-8B}
\newcommand{\qwen}{Qwen1.5-7B}
\newcommand{\llama}{Llama-2-7B}
\newcommand{\mistral}{Mistral-7B}
\newcommand{\stable}{Stablelm-zephyr-3b}

\newcommand{\gemini}{Gemini-1.5-flash}
\newcommand{\gpt}{GPT-3.5-turbo}

\title{On the Effectiveness of LLMs for\\ Manual Test Verifications}

\author{
 \textbf{Myron David Lucena Campos Peixoto \textsuperscript{1}},
 \textbf{Davy de Medeiros Baia \textsuperscript{2}},
 \textbf{Nathalia Nascimento\textsuperscript{3}},
\\
\textbf{Paulo Alencar \textsuperscript{4}},
 \textbf{Baldoino Fonseca\textsuperscript{1}},
 \textbf{M{\'a}rcio Ribeiro \textsuperscript{1}}
\\
\\
 \textsuperscript{1} Federal University of Alagoas (UFAL), Maceio, Alagoas, Brazil \\
 \textsuperscript{2} Federal University of Alagoas (UFAL), Penedo, Alagoas, Brazil \\
 \textsuperscript{3} The Pennsylvania State University, Great Valley, PA, USA \\
 \textsuperscript{4} University of Waterloo, Waterloo, ON, CA
\\
 \small{
   \textbf{Correspondence:} \href{mailto:mdlcp@ic.ufal.br}{mdlcp@ic.ufal.br}
 }
}

\begin{document}
\maketitle
\begin{abstract}
\textbf{Background:} Manual testing is vital for detecting issues missed by automated tests, but specifying accurate verifications is challenging. 
\textbf{Aims:} This study aims to explore the use of Large Language Models (LLMs) to produce verifications for manual tests.
\textbf{Method:} We conducted two independent and complementary exploratory studies. The first study involved using 2 closed-source and 6 open-source LLMs to generate verifications for manual test steps and evaluate their similarity to original verifications. The second study involved recruiting software testing professionals to assess their perception and agreement with the generated verifications compared to the original ones.
\textbf{Results:} The open-source models Mistral-7B and Phi-3-mini-4k demonstrated effectiveness and consistency comparable to closed-source models like Gemini-1.5-flash and GPT-3.5-turbo in generating manual test verifications. However, the agreement level among professional testers was slightly above 40\%, indicating both promise and room for improvement. While some LLM-generated verifications were considered better than the originals, there were also concerns about AI hallucinations, where verifications significantly deviated from expectations.
\textbf{Conclusion:} We contributed by generating a dataset of 37,040 test verifications using 8 different LLMs. Although the models show potential, the relatively modest 40\% agreement level highlights the need for further refinement. Enhancing the accuracy, relevance, and clarity of the generated verifications is crucial to ensure greater reliability in real-world testing scenarios.
\end{abstract}

\section{Introduction}

Manual testing refers to the process of manually executing test cases without the use of automation tools. Testers perform these tests from an end-user perspective to identify defects and ensure that the software functions as expected. Manual testing involves various test techniques, including exploratory, usability, and ad-hoc testing, to uncover issues that automated tests might miss.

Manual testing remains crucial in software engineering despite the rise of automated testing for several reasons. Manual tests (i) leverage human intuition to detect issues that automated tests might miss; (ii) support exploratory testing of the application; (iii)  adapt to changes in the software and deal with rapidly changing projects; and (iv) support tests involving usability, being better suited for subjective criteria like look and feel, ease of use, and user satisfaction.
Specifying manual tests involves detailing the actions that developers need to perform and outlining the expected results to verify after each action. Although specifying manual tests might seem straightforward, involving only actions and verifications, recent studies like \cite{soares2023manual} and \cite{aranda2024catalog} have identified numerous flaws in existing specifications. These studies found actions without verifications or inappropriate verifications, including redundancy, vagueness, and excessive complexity, which can severely undermine the effectiveness of the testing process.

In this context, this paper presents an exploratory study on the effectiveness of large language models (LLMs) to produce verifications for manual tests.  Large language models, trained on diverse and extensive datasets, exhibit remarkable capabilities in understanding and generating human language. These models can be employed to do a variety of tasks \cite{vaswani2017attention}. In the context of manual testing, LLMs can be a promising way to produce textual verifications able to specify properly the expected results of actions performed by testers. In our study, we analyze the effectiveness of six open-source (\mistral, \phimini, \stable, \qwen, \gemma, and \metallama) and two closed-source (\gemini, and \gpt) LLMs. The results suggest that the \mistral and \phimini present the best effectiveness among the open-source LLMs and the closed-source LLMs present effectiveness close to the \mistral and \phimini.  

\section{Natural Language Test Smells}





Manual testing uses natural language to describe test cases. Poorly written test cases can suffer from issues like smells or failures, posing threats to testing activities.

For example, the Unverified Action Smell occurs when an action lacks verification \cite{soares2023manual}. 
Without proper verification, tests may not confirm the application's expected behavior. This can confuse testers about outcomes, such as “What should happen when the message is clicked again?” and “Do additional elements disappear? If not, does the test fail?” Addressing these uncertainties is crucial for clarity and effective test execution.





\section{Study Design}

We evaluate the effectiveness of open and closed-source LLMs in producing manual test verifications. We aim to answer the following research questions: 

\textbf{RQ1. How effective are \textit{open-source} LLMs to produce manual test verifications? }

This research question evaluates the effectiveness of open-source LLMs in producing manual test verifications. To do that, we send several test actions to the open-source LLMs, obtain the verifications produced by these LLMs, and calculate the similarity between the produced verifications and the original ones.  As a result, we have similarity indices indicating how close the produced verifications are to the original verifications. 

\textbf{RQ2. How effective are closed-source LLMs to produce manual test verifications? }

Similarly to the \textbf{RQ1}, the \textbf{RQ2} evaluates the effectiveness of LLMs, but now this research question focuses on closed-source LLMs. As a result,  we expect to reveal the effectiveness of closed-source LLMs in producing manual test verifications that are close to original verifications.

\textbf{RQ3. How effective are open-source LLMs to produce manual test verifications when compared to closed-source LLMs? }

In this research question, we compare the effectiveness of open-source and closed-source LLMs. To do that, we use statistical tests to verify whether there is a statistically significant difference between the similarity indices obtained from each kind of LLM (open and closed-source). As a result, we expect to reveal in which cases it is better to use closed-source LLM instead of open-source LLM and vice-versa. This is important because the use of open-source LLMs may lead to cost reduction in a software project. 

\textbf{RQ4. How do software testing professionals perceive the verifications produced by LLMs? }

In this research question, we evaluate professionals' perceptions of manual test verifications produced by open- or closed-source LLMs. To do that, we recruit six software testing professionals from a large smartphone manufacturer. The name of the company is omitted due to non-disclosure agreements. Answering \textbf{RQ4} is important to better understand whether the software testing professionals find the produced verifications useful to improve the quality of the tests.

\subsection{Dataset} \label{sub:dataset}


Our study uses a dataset sourced from the Ubuntu manual test repository \cite{ubuntu_manual_tests} and cataloged by Aranda et al. \cite{aranda2024catalog}. This dataset includes 973 manual test cases with a total of 6,598 test steps. Each test step is structured as a tuple of actions and verifications (expected results), though some steps are missing actions (47) or verifications (1968, approximately 29.83\%). The Unverified Action Smell is one of the test smells identified by Aranda et al. For our experiment, we focused on the 4,630 test steps that contain both actions and verifications.

Table \ref{tab:dataset} illustrates an example of a manual test from the Ubuntu OS, showing the tuple of actions and verifications for each step.

\begin{table}[!htb]
\centering
\caption{Example of a Manual Test from the Ubuntu OS.}
\begin{tabular}{|ccc|}
\hline
\multicolumn{3}{|l|}{\textbf{Manual Test ID: 1412}}     \\ \hline
\multicolumn{3}{|l|}{\begin{tabular}[c]{@{}l@{}}Name: Unity GSetting Migration\\ Precondition: This testcase is intended to...\end{tabular}}                                                                           \\ \hline
\multicolumn{1}{|c|}{\textbf{\#Step}} & \multicolumn{1}{c|}{\textbf{Actions}}                                                                 & \textbf{Verifications}                                                \\ \hline
\multicolumn{1}{|c|}{1}                & \multicolumn{1}{c|}{\begin{tabular}[c]{@{}c@{}}Open the dash and \\ launch `appearance'\end{tabular}} & \begin{tabular}[c]{@{}c@{}}Appearance \\ applet launches\end{tabular} \\ \hline
\multicolumn{1}{|c|}{2}                & \multicolumn{1}{c|}{\begin{tabular}[c]{@{}c@{}}Open the dash and \\ launch `keyboard'\end{tabular}}   & \begin{tabular}[c]{@{}c@{}}Keyboard \\ applet launches\end{tabular}   \\ \hline
\multicolumn{1}{|c|}{...}              & \multicolumn{1}{c|}{...}                                                                              & ...                                                                   \\ \hline
\end{tabular}
\label{tab:dataset}
\end{table}

\subsection{Process}

\begin{figure}[!htb]
\centering
\includegraphics[width=0.49\textwidth]{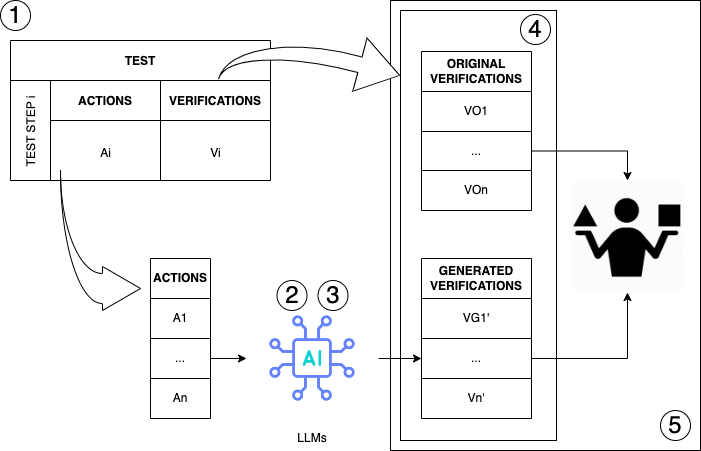}
\caption{Study Design.}
\label{fig:steps}
\end{figure}

Figure \ref{fig:steps} illustrates the flow of our experimental process. Our study consists of five steps:

\textbf{S1. Test Step Selection:} We start by processing the dataset and selecting test steps that have both actions and verifications, as described in subsection \ref{sub:dataset}. 

\textbf{S2. Closed-Source LLM Verifications:} For generating verifications, we use selected closed-source models (\gemini \, and \gpt), chosen for their popularity and reliability. The applied prompt is:

\begin{quoting}
\texttt{Consider a manual test which has a precondition and a list of steps with actions and verifications. Given the precondition \{precondition\}, complete a test step generating the reaction for the following action: \{action\}. Only generate the verification in one line and return it in raw text.}
\end{quoting}

Here, \{precondition\} refers to the test precondition or header, and \{action\} refers to the actions of the test step.

\textbf{S3. Open-Source LLM Verifications:} We also use  open-source models (\gemma, \phimini, \metallama, \qwen, \llama, \mistral, and \stable)  
to generate verifications. These models were selected for their trending status on Hugging Face and diverse architectures. The actions were input into these models using the same prompt structure as the closed-source ones.

As a result of steps 2 and 3, where we generated a verification for each one of the 4630 test steps with the 8 models, we created a dataset of 37040 generated test verifications.

\textbf{S4. Similarity Analysis:} A similarity function is applied to compare the 37040 generated verifications with the original ones and to select samples for the experiment with the software testing professionals, as described in subsection \ref{sub:similarity}. This step involves using similarity-based sampling to ensure diverse representation and reduce bias.

\textbf{S5. Evaluation by Professional Testers:} Finally, professional testers evaluate a subset of the generated verifications. Detailed information about this validation experiment with the testers is provided in subsection \ref{sub:participants}.

\subsection{Similarity Technique} \label{sub:similarity}

We used the Semantic Textual Similarity (STS) technique \cite{reimers-2019-sentence-bert} with the cosine similarity metric to evaluate the verification generated by the LLMs. This technique compares the generated verification text with the original verification text, where higher similarity scores indicate greater semantic similarity.

These similarity results are used at two critical points in our experiment. First, to evaluate the effectiveness of the LLMs by assessing how closely the generated verifications match the original ones.  Second, to select samples for testing with professional testers. For each model, we grouped the test steps into five similarity ranges: 0-0.2, 0.2-0.4, 0.4-0.6, 0.6-0.8, and 0.8-1, aiming to remove bias by including samples with low similarity scores that still hold meaningful content.

For each of the 8 LLM models, we randomly selected 3 samples from each similarity group, resulting in a total of 120 samples (8 models * 5 groups * 3 samples). This process was repeated for each of the 6 participants, resulting in a total of 720 samples to be manually evaluated. This ensured a diverse and balanced sample selection for the experiment. 

\subsection{Participants} \label{sub:participants}


For this study, we recruited six software testers to evaluate the generated verifications. Each participant received 120 tuples, with each tuple containing the original verification and the corresponding generated verification.

Participants were asked, ``Do you agree with the generated verification?" using a Likert scale from "Strongly Disagree" to "Strongly Agree." This quantitative method assessed whether participants perceived the generated verification as similar to the original.

This study investigates the (dis)agreement among participants on the manual tests generated by the 2 closed-source and 6 open-source models. In this step, we also explore whether certain factors, such as the length of the tests, influence the similarity or dissimilarity of software testers' perceptions of the generated tests.

\section{Results}

This section describes and discusses the research questions analyzed in our study.

\subsection*{RQ1) How effective are \textit{open-source} LLMs to produce manual test verifications? }

\begin{figure*}[!t]
\centering
\vspace{-4ex}
\includegraphics[width=\textwidth]{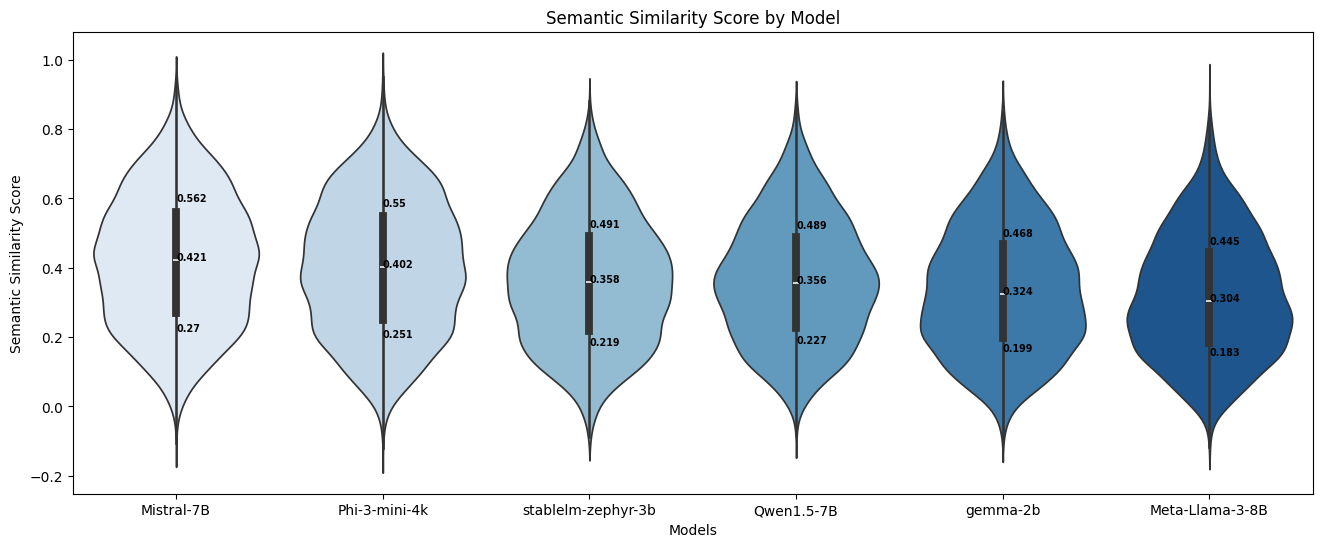}
\caption{Open-source Models.}
\label{fig:results_rq1}
\end{figure*}

In \textbf{RQ1}, we evaluate the effectiveness of open-source LLMs. Figure~\ref{fig:results_rq1} shows violin plots representing the similarity distribution between the actual verifications and the ones produced by seven different open-source models: \gemma, \phimini, \metallama, \qwen, \llama, \mistral, and \stable. Each violin plot represents the distribution of similarity values for each model, with key statistical measures (median and interquartile range - IQR) highlighted. The median represents the middle value of the similarity scores, meaning half of the scores are above this value and half are below. A higher median indicates that, overall, the model tends to produce higher similarity scores, which may suggest better performance in terms of semantic similarity. The IQR measures the spread of the middle 50\% of the data, calculated as the difference between the 75th percentile (Q3) and the 25th percentile (Q1). A low IQR indicates that the similarity scores are closely clustered around the median, suggesting high consistency and low variability. This means the model reliably produces similar scores across different inputs. A high IQR indicates a wider spread of similarity scores, suggesting greater variability. This could mean the model is more sensitive to different inputs and produces a wider range of similarity scores, which may be desirable in certain contexts but could also indicate less predictability.

\textbf{\mistral.} The distribution is relatively wide, with a peak around the median. The median similarity score is around 0.42, indicating a moderate level of similarity between the actual verifications and the verifications produced by the models. In software testing, a cosine similarity of 0.42 between two verification vectors might indicate that the verifications share some common terms or topics. Fortunately, reaching a high similarity is unnecessary to obtain a reasonable verification since it just needs to be coherent with the actions described in the tests. We also observe that the first and third quartiles are 0.27 and 0.56, respectively, resulting in an IQR of 0.29. This indicates a low variability of the similarity scores, suggesting that the model is consistent in the produced verifications.  

\textbf{\phimini.} Similarly to \mistral, the model \phimini \hspace{0.1cm} presents a peak near the median and low variability. The median similarity score is around 0.402, with an IQR from 0.251 to 0.55 (IQR = 0.3), suggesting that \phimini can consistently produce reasonable verifications.

\textbf{\stable.}  This model presents a median similarity score slightly lower than \mistral \hspace{0.1cm}and \phimini. \stable \hspace{0.1cm}
presents a median of around 0.36, indicating that these models can produce reasonable verifications. Notice also that the \stable \hspace{0.1cm} presents an IQR (0.27) slightly greater than \mistral and \phimini, indicating that \stable \hspace{0.1cm} presents a slightly better consistency in the production of verifications than \mistral \hspace{0.1cm} and \phimini. 

\textbf{\qwen.} This model presents a median similarity score equal to \stable, obtaining a median of around 0.36. Regarding the IQR, \qwen \hspace{0.1cm} presents a value of 0.26, indicating a consistency in the production of the verifications similar to \stable. Thus, both the models \stable \hspace{0.1cm}and \qwen \hspace{0.1cm} present effectiveness and consistency very similar in the production of verifications. 

\textbf{\gemma and \metallama.} These models present median similarity scores varying from 0.3 (\metallama) to 0.32 (\gemma). Although these scores are slightly lower than the previously analyzed models, we observe that the models \gemma \hspace{0.1cm}and \metallama \hspace{0.1cm}still produce reasonable verifications. Regarding the IQR, these models present values close to those previously analyzed models. This indicates that the models \gemma \hspace{0.1cm}and \metallama \hspace{0.1cm}present consistency in the produced verifications. 


\textbf{Summary.} The models \mistral \hspace{0.1cm}and \phimini \hspace{0.1cm}present the best effectiveness and consistency in the production of verifications. 

\subsection*{RQ2) How effective are closed-source LLMs to produce manual test verifications? }

\begin{figure}[!t]
\centering
\vspace{-4ex}
\includegraphics[width=0.4\textwidth]{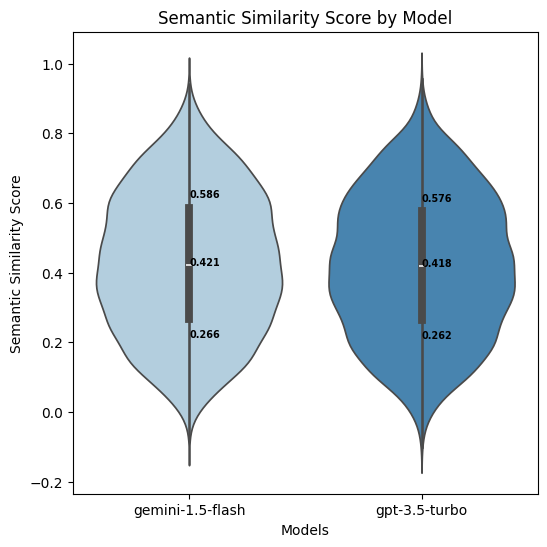}
\caption{Closed-source Models.}
\label{fig:results_rq2}
\end{figure}

In \textbf{RQ2}, we evaluate the effectiveness of closed-source LLMs. Figure~\ref{fig:results_rq2} shows violin plots representing the similarity distribution between the actual verifications and the ones produced by two different closed-source models: \gemini \, and \gpt. Each violin plot represents the distribution of similarity values for each model. 

\textbf{\gemini \, and \gpt.} Both these models present a distribution relatively wide, with a peak around the median. The median similarity score is around 0.42, indicating a moderate level of similarity between the actual verifications and the verifications produced by the models. Also, we observe that both present an IQR of 0.32. This indicates a low variability of the similarity scores, suggesting that the model is consistent in the verifications produced.  

Compared with the open-source models, we observe that the models \gemini \, and \gpt \hspace{0.1cm}present a median similarity score equal to the \mistral. Regarding the IQR, \gemini \, and \gpt \;present a similarity scores variability slightly greater than the \mistral. 

\textbf{Summary.} The results indicate that the open-source models are as effective as closed-source models.

\subsection*{RQ3) How effective are open-source LLMs to produce manual test verifications when compared to closed-source LLMs?}

\begin{figure*}[!t]
\centering
\vspace{-4ex}
\includegraphics[width=1\textwidth]{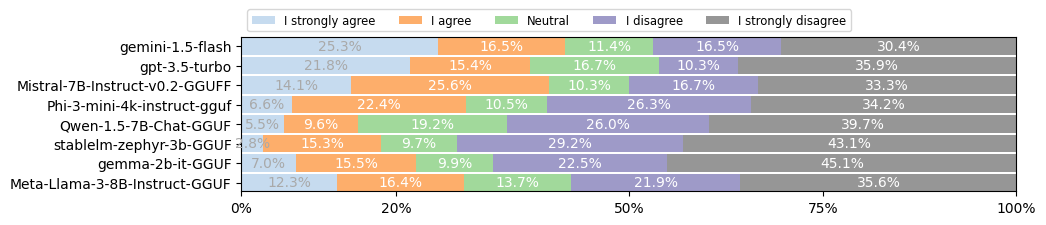}
\caption{Professional Responses.}
\label{fig:results_rq3}
\end{figure*}

In this research question, we analyze the professionals' perceptions on the verifications produced by LLMs. Figure \ref{fig:results_rq3} presents the distribution of responses for different LLMs, categorized by their source (closed or open). Also this, shows the percentage of responses for each agreement level across different models, including both closed source (gemini-1.5-flash, gpt-3.5-turbo) and open source (TheBloke/Mistral-7B-Instruct-v0.2-GGUFF, microsoft/Phi-3-mini-4k-instruct-gguf, Qwen/Qwen-1.5-7B-Chat-GGUF, TheBloke/stablelm-zephyr-3b-GGUF, lmstudio-ai/gemma-2b-it-GGUF, QuantFactory/Meta-Llama-3-8B-Instruct-GGUF). The open-source models received the most negative feedback and the closed-source models performed relatively better in terms of generating acceptable verification. These findings highlight the importance of continuous improvement and customization of LLMs for specific tasks like test case verification. Moreover, they underscore the need for further research and development to enhance the reliability and accuracy of verification generation processes. The figure clearly illustrates the varying levels of agreement among test analysts regarding the quality of generated verification by different LLMs. While some models show promise, there is significant room for improvement, especially in ensuring that automated verification meets the expectations and standards of human test analysts.



\textbf{RQ4) How do software testing professionals perceive the verifications produced by LLMs? }


In Figure \ref{fig:results_rq3} 
it is possible to notice a predominantly negative perception of the LLM generated  verifications among test analysts. The high percentage of "Strongly Disagree" and "Disagree" responses highlights significant skepticism regarding the current effectiveness of LLMs in generating reliable test case verifications. The relatively lower percentages of "Agree" and "Strongly Agree" suggest that while there is some recognition of the utility of LLMs, it is limited and not widespread. This  portion of the respondents who recognize the potential utility of these models point to an area for potential improvement and future research.

These results suggest a need for further development and refinement of LLMs to better meet the needs and expectations of test analysts. Enhancements in the accuracy, relevance, and clarity of generated verifications could improve acceptance and satisfaction levels among users.

\section{Discussion}

In this section, we discuss the main results. 

As shown in RQ1 and RQ3 results, \mistral \, and \phimini models presents more consistency and effectiveness in producing verifications. Looking at the specific verifications generated from these models, we notice that the original verifications of the test steps where the generated verifications are more acceptable (in terms of developers' perceptions) commonly involve short and direct sentences. The sample below have a generated verification from \mistral \, with a high semantic score of 0.93 but labeled in the "Strongly Agree" group.
Action:
\begin{quoting}
    \texttt{Write "ping" into the "Reply" input field, press "Send" and drag up to the top of the timeline}
\end{quoting}
\noindent Original Verification:
\begin{quoting}
    \texttt{A message containing "ping" is shown at the top of the timeline}
\end{quoting}
Generated Verification:
\begin{quoting}
    \texttt{The "ping" message appears at the top of the timeline.}
\end{quoting}

Although developers' perceptions typically align with high similarity scores, discrepancies can occur. For instance, a generated verification with a similarity score of 0.34 was still labeled as "Strongly Agree."

Action:
\begin{quoting}
    \texttt{Move the Volume control (up/down)}
\end{quoting}
Original Verification:
\begin{quoting}
    \texttt{Did you feel the change of volume?}
\end{quoting}
Generated Verification made by \phimini \, model:
\begin{quoting}
    \texttt{Verify that the volume level changes as expected when adjusting the controls.}
\end{quoting}

Similarly, there are cases where samples with high similarity scores are labeled as "Strongly Disagree," such as a verification generated by the \qwen , model.

Action:
\begin{quoting}
    \texttt{Note the state of the 'Erase disk and install FAMILY' radio button}
\end{quoting}
Original Verification:
\begin{quoting}
    \texttt{The 'Erase disk and install FAMILY' radio button is selected}
\end{quoting}
Generated Verification made by \phimini \, model:
\begin{quoting}
    \texttt{Verify that the radio button is disabled}
\end{quoting}

\textbf{Closed-source LLM.} 


As shown in Figure \ref{fig:results_rq3}, the \gemini , model outperformed \gpt slightly, with more "Strongly Agree" and fewer "Strongly Disagree" responses, both close in similarity scores and professional perceptions.

Both open-source and closed-source models perform better with short, direct commands in tests. 


We observe the same behavior previously  related to open source models in cases where we have a low semantic similarity score but the generated sentence was cataloged in the "Strongly Agree" group. 


The same occurs in the other direction, with the generated sentence having a high similarity score belonging to the "Strongly Disagree" group. 


\textbf{Open x Closed-source LLM.} 
As seen in Figure \ref{fig:results_rq3}, the closed-source models obtained some advantage over the open source ones. This can be exemplified by the sample below, where both closed-source models obtained 0.81 in the semantic similarity score while all open source models obtained less than 0.4, with the exception of the \phimini \, model, which obtained 0.62 in the similarity score.

The action in this case is more complex and if we look at the analyzed test step, we can identify more than one smell based on the catalogs of \cite{hauptmann2013hunting} and \cite{soares2023manual}.

Action:
\begin{quoting}
    \texttt{If there is only one hard disk, skip to step 12 (On the 'Where are you?' screen...). Otherwise, on the 'Installation type' screen verify that the drive selected on the Select drive list corresponds to the drive on the chart (e.g /dev/sda)}
\end{quoting}

A contributing factor for the difference in performance in the analyzed models is that each one of them have considerable particularities. The \phimini \, model , for example, has 3.7B parameters while  \mistral has 7B. Further, the closed source models are pre-trained with trillions of parameters, which makes the comparison unfair. Despite of that, the results associated with the specific task of generating verifications indicate that there are no super gains of the closed-source in comparison with the open source models.

\textbf{Professionals' Perceptions.}
During the evaluation stage of this study, the volunteers made some pertinent comments about the evaluated samples. For instance, as a positive perception, some generated verifications were considered better than the original ones. In contrast, as a negative perception, they noticed a few samples of AI hallucination, with verifications completely different from the expected.

\section{Implications}

This section presents the implications of our study for researchers and practitioners.

The discussed results highlight key concerns. In the analyzed samples, many low-acceptance verifications could improve with prompt refinement. Lack of context in prompts can lead to verifications that initially seem suitable but yield completely different results.

Test actions are crucial in generating verifications. Flawed tests can produce poor results, adding more issues.

The findings show no significant differences between open-source and closed-source models. The critical factor appears to be the number of parameters. As shown in Figure \ref{fig:results_rq3}, models with fewer parameters, such as \gemma , and \stable, perform worst.

Understanding these concerns can guide future research and practical applications, aiding model selection and experiment design. All models used are widely available, facilitating replication and new studies.

We also produced a dataset with 37,040 test verifications generated by 8 different LLMs. This dataset includes complete data from Ubuntu's original tests in a structured database, enabling detailed queries for future experiments.

\section{Related Work}
Previous works already catalogued a set of test smells that occurs on natural language tests. \cite{hauptmann2013hunting} was the precursor, defining smells for manual tests in natural language as Natural Language Test Smells (NLTS). He defined seven types of smells. 


A few years later, \cite{Rajkovic6382} proposes some metrics to analyse software requirements in a more generalized way. Despite involving testing, it was not the main focus of the work. The work also present a tool to automatically detect bad smells in software requirements based on the proposed metrics.

Ten years after \cite{hauptmann2013hunting}, \cite{soares2023manual} presented six new smells and updated the set. In both studies, the smells are well defined and we are presented with rules to identify them. \cite{soares2023manual} also present a NLP-based tool that automaticaly identifies smells on manual tests.

More recently, in April 2024, \cite{wang2024software} presents a comprehensive review of the utilization of pre-trained large language models (LLMs) in software testing. This survey analyzes 102 relevant studies, exploring the use of LLMs from both software testing and LLM perspectives. Key areas of focus include test case preparation and program repair, with detailed discussions on common LLMs used, types of prompt engineering, and associated techniques. The paper also identifies key challenges and potential opportunities, providing a roadmap for future research in this domain. 

Also, in 2024, \cite{aranda2024catalog} proposes a template to represent a natural language test and ways to treat the natural language test smells, which he calls "transformations." In total, the study presents a set of seven transformations for seven different smells, such as "Extract Action" transformation for the smell "Misplaced Action." \cite{aranda2024catalog} also present a tool to make the transformations of the smells automatically using NLP.


In the LLM field, there are a few studies that addresses the use of LLMs to augmenting software testing methods. For example, \cite{siddiq2024using} presented an empirical work on the use of LLMs to generate unit tests for the Java programming language. \cite{yuan2023no} develops a similar approach, evaluating the capacities of OpenAI's ChatGPT\index{net}\footnote{Available in \url{https://chatgpt.com/} } for test generation and proposing a more powerful fine-tuned model. However, none of this studies focuses on test smells, manual tests or more specifically, natural language test smells.

\section{Threats to Validity}

There are potential threats to validity in our experiment. \textbf{Subjectivity in evaluation}: Evaluating manual tests by specialists can be subjective. To mitigate this, we used the original tests as a baseline. \textbf{Multiple Valid Descriptions and Test Smells}: In manual testing, it is possible to have multiple valid descriptions for the same expected results. Additionally, the presence of test smells in the test repository, as cataloged by Aranda et al. \cite{aranda2024catalog}, can generate ambiguity. To address these issues, we included manual verification by test specialists and incorporated samples from different similarity score groups. \textbf{Human-generated baseline}: We used human-generated tests as the ideal baseline. However, these tests are not error-free. The specialists noted that, in some cases, the generated tests surpassed the human-generated ones in quality. \textbf{Minimizing disagreements and errors}: Subjectivity and potential technical misunderstandings could lead to disagreements and errors. We minimized this by involving six specialists in the evaluation process. 

\section{Conclusion and Future Work}

This paper is part of an ongoing effort to use LLMs to improve the quality of manual tests. At this first stage, we performed two independent and complementary exploratory studies to evaluate the efficacy of LLMs in generating verifications for manual tests.

We evaluated 2 closed-source models and 6 open-source models. In the first experiment, we used the original tests as a baseline and performed a similarity evaluation, resulting in a dataset of 37,040 generated verifications. Each entry included the actual verification test, the LLM model used, and the similarity score. Our similarity evaluation between actual and generated verifications revealed that the open-source models Mistral-7B and Phi-3-mini-4k, along with the closed-source models Gemini-1.5-flash and GPT-3.5-turbo, demonstrated the best effectiveness and consistency.

In the second experiment, we recruited 6 software testers to evaluate the generated tests compared to the original ones. Despite including tuples with low similarity scores for evaluation, we achieved similar results: the open-source model Mistral-7B and the two closed-source models, Gemini-1.5-flash and GPT-3.5-turbo, received the highest acceptance scores. In the case of the two closed models, the sum of ``Strongly Agree," ``Agree," and ``Neutral" responses exceeded 50\%. Notably, the models with fewer parameters had the highest disagreement levels.

An agreement level exceeding 40\% suggests some promise in this approach, but it also indicates that further research is needed before LLMs can be reliably used to generate manual test verifications in practical settings. \textbf{In future work}, we plan to explore the use of multiple LLMs in ensemble or voting setups, as well as architectures where one LLM evaluates or agrees with another, to improve the accuracy of generated verifications. Additional experiments will assess the confidence of LLMs in generating manual test verifications, exploring different models, contexts, and larger datasets to better understand their capabilities and limitations. We also aim to investigate the efficacy of LLMs in addressing test smells, such as ambiguous tests, misplaced verifications, and unverified actions. For example, in the Ubuntu repository used in this study, 29.8\% of test steps have the Unverified Action Smell. Generating verifications for these steps could improve the repository's completeness and quality. However, the risk of incorrect verifications may be higher than missing ones. Therefore, it's important to investigate whether generated verifications might increase undetected faults and other test smells.


\section{Limitations}
This study has some key limitations. \textbf{Subjectivity in Evaluation}: Despite our efforts to mitigate subjectivity by using original tests as a baseline, evaluating manual tests by specialists remains inherently subjective. \textbf{Scope of Test Smells}: We focused primarily on addressing the Unverified Action Smell, but we are aware of the presence of other test smells in the repository. Despite this, we did not remove these other smells, which may have influenced the quality of the original tests. \textbf{Generalizability of Results}: While we used both automated and manual verification methods, including similarity analysis for 37,040 generated test verifications and a manual review of 720 samples by six specialists, our sample is limited to one system (Ubuntu OS) and relies solely on tests from the Ubuntu Quality Team. This limitation may affect the applicability of our findings to other systems or testing environments. \textbf{Model Specificity}: The LLMs used were selected based on their popularity and availability. Different models or future versions may yield different results. \textbf{Tester Expertise}: The evaluations by software testing professionals are influenced by their individual expertise and experience.


\bibliography{acl-latex.bib}

\end{document}